\documentclass[10pt,letterpaper]{article}

\newcommand{\expect}[1]{\langle#1\rangle}

\def\TR{T_{{\rm R}}}
\def\frep{f_{{\rm rep}}}  \def\fceo{f_{0}}
\def\bv{{\bf v}} \def\bS{{\bf S}}
\def\sA{{\sf A}} \def\sI{{\sf I}} \def\sD{{\sf D}}
\def\om{\varpi}
\def\Dw{\Delta w} \def\Do{\Delta\varpi} \def\Dt{\Delta\tau}  \def\Dth{\Delta\theta} \def\Dg{\Delta g}
\def\geq{g_{\rm eq}}  \def\weq{w_{\rm eq}}
\def\omeq{\om_{\rm eq}}
\def\tph{\tau_{\rm ph}}
\def\Npheq{N_{\rm ph,eq}} \def \Ntwoeq{N_{2, \rm eq}}

\usepackage{opex3}
\usepackage{amsmath}
\usepackage{graphicx}

\begin{document}
\title{The quantum-limited comb lineshape of a mode-locked laser: 
Fundamental limits on frequency uncertainty}
\author{J. K. Wahlstrand,$^1$ J. T. Willits,$^{1,2}$ C. R. Menyuk,$^3$ and S. T. Cundiff$^1$}
\address{$^1$JILA, National Institute of Standards and Technology and the University of Colorado, Boulder, CO 80309-0440 \\
$^2$Department of Electrical and Computer Engineering, University of Colorado, Boulder, CO 80309 \\
$^3$Computer Science and Electrical Engineering Department, University of Maryland Baltimore County, 1000 Hilltop Circle, Baltimore, MD 21250
}

\begin{abstract}
We calculate the quantum-limited shape of the comb lines from a mode-locked 
Ti:sapphire laser using experimentally-derived parameters for the linear
response of the laser to perturbations.  The free-running width of the comb lines is found across the laser spectrum.
By modeling the effect of a simple feedback 
loop, we calculate the spectrum of the residual phase noise in terms of the quantum noise and the feedback parameters.    
Finally, we calculate the frequency uncertainty in an optical frequency measurement if the limiting factor is quantum noise
in the detection of the optical heterodyne beat.
\end{abstract}



\section{Introduction}
When light passes though any gain medium, it will acquire noise through
spontaneous emission (SE) \cite{loudon_quantum_2000}. The
SE noise sets a fundamental limit on the linewidth of a laser.  For a continuous
wave laser, this Schawlow-Townes limit is due to phase jitter \cite{schawlow_infrared_1958}.
The situation with modelocked lasers is
more complicated.  A mode-locked laser produces a train of pulses, regularly
spaced in time, and the frequency spectrum is a series of narrow comb lines.  
The width of each comb line depends on the timing and phase jitter of the pulse train.
The laser dynamics are characterized by four pulse parameters:  the pulse energy, the central or carrier frequency, the central pulse time, the phase, and in addition a fifth parameter:  the round-trip gain. 
In all mode-locked lasers, strong nonlinearities couple amplitude and frequency fluctuations into timing and phase jitter.
The quantum-limited noise properties of mode-locked lasers were
first treated by Haus and Mecozzi in \cite{haus_noise_1993}, hereafter referred
to as HM.  Until recently, subsequent work focused mainly on quantum-limited
timing jitter.  With the development of femtosecond frequency combs
\cite{cundiff_2003}, the more general problem of timing and phase jitter has received more attention
\cite{kaertner_few-cycle_2004,paschotta_noise_2004,paschotta_noise_2004-1,newbury_theory_2005,paschotta_optical_2006,matos_carrier-envelope_2006,ablowitz_2006,newbury_low-noise_2007}.
While HM calculated the timing and phase jitter rather than the comb 
linewidths, that paper contains almost all of the information needed to calculate the linewidths.
More recent work has included making extensions
to the HM approach to determine the linewidths when bounded contributions to the 
timing and phase jitter can be ignored \cite{kaertner_few-cycle_2004}, developing
more elaborate models of pulse propagation in the laser
\cite{paschotta_noise_2004,paschotta_noise_2004-1}, including technical noise contributions \cite{newbury_theory_2005, newbury_low-noise_2007}, and including the effects of 
gain dynamics \cite{matos_carrier-envelope_2006}.
A major unknown in all theoretical efforts to date was the strength of couplings between pulse parameters.  Theoretical predictions carry large uncertainties and depend on the details of the laser design, including dispersion management.  Recently, we quantitatively measured the linear response of the pulse energy,
the central frequency, the round-trip gain \cite{menyuk_pulse_2007}, the timing 
and the phase \cite{wahlstrand_quantitative_2007} of a mode-locked Ti:sapphire
laser.  The motivation for that work was to quantitatively predict
the SE-limited noise properties of the laser, the subject of this paper.

When considering the shape of the comb lines of a mode-locked laser, 
there is a fundamental distinction between the case of a free-running laser 
and the case of a laser locked to an external 
oscillator.  In the former case, when the noise source is SE noise or any other white noise source, the central 
time and the phase of the mode-locked pulse undergo a random walk, and the 
comb lines have a stationary shape.  In the latter case, again 
assuming white noise, the central pulse time and the phase are 
bounded, and there is no stationary line shape; the measured line width is 
inversely proportional to the measurement time.  However, the phase noise 
spectrum of each comb line is stationary, and that spectrum --- not the frequency spectrum --- is physically 
meaningful.  Any clock or frequency measurement system consists of an 
oscillator and a counter \cite{nist_1337}.  The clock's performance depends on the 
frequency noise of the oscillator and the phase noise of the counter. 
Virtually all theoretical calculations to date of the noise properties of 
passively mode-locked lasers have focused on the frequency spectrum of a 
free-running laser, but in modern time and frequency metrology 
applications, these lasers are part of the counting system \cite{cundiff_2003},
so it is their phase noise spectrum after they are 
locked to an oscillator that is important.
Here, we calculate the SE-limited noise properties of a Ti:sapphire laser,
comparing the comb line shape of the free-running laser and the residual
phase noise of comb lines of the phase-locked laser,
using experimentally-derived parameters for the linear response of the laser to noise perturbations.

\section{Linear response and noise drivers}
The laser is characterized by the pulse energy $w=\weq+\Dw$, the gain
$g=\geq+\Dg$, the central frequency $\om=\omeq+\Do$, the central pulse time
$\tau=\tau_{\rm eq}+\Dt$, and the phase $\theta=\theta_{\rm eq}+\Dth$, where
$x_{{\rm eq}}$ denotes the equilibrium value of each quantity and $\Delta x$ the
fluctuation about equilibrium.  Defining $\TR$ as the round trip time, we have
the repetition rate $\frep=1/\TR$.  Without loss of generality, we may choose $\tau_{\rm eq}=0$ and
$\theta_{\rm eq}=2\pi f_0 \TR$, where $f_0$ is the equilibrium carrier-envelope
offset frequency.  We define the phase
with respect to a common reference rather than the pulse envelope.  Consequently,
$\theta$ is related to the carrier-envelope phase slip
$\Delta\theta_{\rm ce}$ by $\Dth=-\Delta\theta_{\rm ce}+\om \Dt$.
The linear response of the laser to perturbations is \cite{haus_noise_1993}
\begin{equation}
\frac{{\mathrm d}\bv}{{\mathrm d}T} = -\sA \cdot \bv + \bS,
\end{equation}
where $\bv=(\Dg,\Dw,\Do,\Dt,\Dth)^t$ ($t$ denotes the transpose), $\bS$ is a vector of noise sources for
each parameter, and $\sA$ is a matrix of coefficients that describe the linear
response of each parameter to changes in itself or the others.  Once $\sA$ and
$\bS$ are known, one can calculate the timing and phase jitter
\cite{haus_noise_1993}, which leads to the optical frequency comb line shapes
\cite{kaertner_few-cycle_2004}.  The noise sources perturb all five parameters, and also add energy into a continuum of dispersive modes,
but these modes must be damped in order for a passively mode-locked laser to be 
stable. Consequently, the noise contributions to those modes do not affect the
linewidths.

To calculate the frequency or phase noise spectrum, we use the formalism of stochastic differential
equations \cite{wai_polarization_1996}. The second-order moment equations for
the laser parameters are
\begin{equation}
\frac{d\langle \Delta v_i \Delta v_j \rangle}{dT} = -\sum_{k,l}(A_{ik}+A_{jl})\langle \Delta v_k \Delta v_l \rangle + D_{ij},
\label{moments}
\end{equation}
where $v_i$ is one of the dynamical variables and
$\langle S_i(T) S_j(T') \rangle = D_{ij} \delta(T-T')$ relates $\bS$ to a matrix $\sD$ of noise drivers.
For the 5 dynamical parameters introduced above, there are thus 15
differential equations describing the response of the second-order moments
to the noise sources.
The quantum noise drivers $D_{ij}$ were found using a generalized perturbation
theory for Gaussian pulses in a dispersion managed soliton laser
\cite{menyuk_pulse_2007}.  Defining the cavity lifetime $\tau_{\mathrm{ph}}$, the fluorescence lifetime $\tau_{\mathrm{f}}$, the equilibrium photon number $\Npheq$, the equilibrium upper state population $\Ntwoeq$, and the pulse width $t_{\mathrm{p}}$, they are
\begin{equation}
\begin{aligned}
D_{ww} =& 2 \weq^2 \frac{1}{\tph} \frac{\hbar \omeq}{\weq},& D_{wg} = D_{gw} =& \weq \geq \frac{\Npheq}{\Ntwoeq}\frac{1}{\tph}\frac{\hbar \omeq}{\weq}, \\
D_{gg} =& 2\geq^2\left[\frac{\tph}{\tau_f}\frac{\Npheq}{\Ntwoeq}+\left(\frac{\Npheq}{\Ntwoeq}\right)^2\right]\frac{1}{\tph}\frac{\hbar \omeq}{\weq}, & D_{\om\om} =& t_p^{-2} \frac{1}{\tph}\frac{\hbar \omeq}{\weq}, \\
D_{\theta\theta} = & 2 F(t_p) \frac{1}{\tph}\frac{\hbar \omeq}{\weq}, & D_{\tau\tau} = &t_p^2\frac{1}{\tph}\frac{\hbar \omeq}{\weq},
\label{drivers}
\end{aligned}
\end{equation}
where $F(t_p) = C_1^2(t_p)-C_1(t_p)C_2(t_p)+(3/4)C_2^2(t_p)$, 
$C_1(t_p)=(1+3s/t_p^4)/(1+5s/t_p^4)$, $C_2(t_p)=(1+s/t_p^4)/(1+5s/t_p^4)$, and
$s$ is a dispersion management parameter.  The results above are similar to previously
used quantities derived from ordinary soliton perturbation theory
\cite{haus_noise_1993}. While most of the drivers differ by small factors, 
the energy and gain drivers are identical.

For the laser studied in
\cite{menyuk_pulse_2007} and \cite{wahlstrand_quantitative_2007},
$\weq=55$ nJ, $\geq=0.054$, $\omeq=2.3\times 10^{15}$ rad/s, $\tph=0.1$ $\mu$s,
$\Ntwoeq/\Npheq \approx 13$, and $s/t_p=0.05$, which leads to $F(t_p)=0.60$.  We
use the experimentally-measured values from \cite{menyuk_pulse_2007} and \cite{wahlstrand_quantitative_2007} for the elements of $\sA$.
One could solve Eq.~(\ref{moments}) in the time domain numerically, but there
are practical difficulties arising from the wide range of time scales involved
in the dynamics.  We instead found the moments in the frequency domain, using
the formal solution
\begin{equation}
\begin{aligned}
\expect{\Delta\tilde{v}_m \Delta\tilde{v}_l^*} (\omega) =& (-i\omega \sI-\sA)^{-1} \cdot \sD \cdot [(i\omega \sI-\sA)^{-1}]^t, 
\end{aligned}
\label{freq_domain_sols}
\end{equation}
where $\sI$ is the identity matrix.  We then used the
inverse Fourier transform to bring this solution into the time domain symbolically.

\section{Free-running lineshape}
\begin{figure}
\center{\includegraphics[width=12cm]{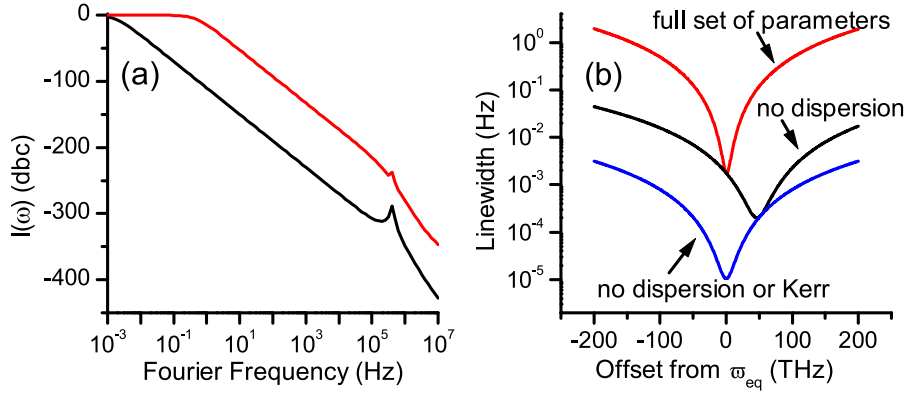}}
\caption{The calculated free-running comb lineshape.
(a)  Power spectrum $I_n(\omega)$ for comb lines near the central frequency (black) and 200 THz from the central frequency (red).  (b) Linewidth $\gamma$ as a function of $\Omega_n$, the offset from $\omeq$.  The red curve is the
result with the full set of $\sA$ coefficients measured in \cite{menyuk_pulse_2007} and \cite{wahlstrand_quantitative_2007}.  The
black curve is the result when timing jitter caused by dispersion is neglected
(by setting $A_{\tau\om}=0$).
The blue curve is the result when, in addition, timing and phase jitter driven by intensity fluctuations is neglected
(by setting $A_{\tau w}=0$ and $A_{\theta w}=0$).
}
\label{linewidth}
\end{figure}

Taking into account only the effects of timing and phase jitter (amplitude and
central frequency noise both contribute to the linewidth, but are in practice
negligible, c.f.~\cite{schawlow_infrared_1958}), the power spectral density of the $n^{\mathrm{th}}$
comb line is
\begin{equation}
I_n(\omega) \propto \int_0^\infty dT \cos(\omega T) \exp \left\{-\frac{1}{2}\Omega_n^2 \langle[\Dt(T)]^2\rangle+\Omega_n\langle\Dt(T)\Dth(T)\rangle-\frac{1}{2}\langle[\Dth(T)]^2\rangle\right\},
\label{pow_spect_dens}
\end{equation}
where $\omega$ is the Fourier frequency with
respect to the center of the comb line $\omega_n$, and $\Omega_n = \omega_n - \omeq$.  See Fig.~\ref{linewidth} for plots of the free-running noise properties.
With no feedback on $\frep$ and $\fceo$, the second-order moments involving only $\tau$ and $\theta$ do 
not evolve to equilibrium values at large $T$, but rather grow with time.  Each can be expressed as $\gamma T + \beta(T)$, where $\gamma$ is a constant and $\beta(T)\rightarrow 0$ as $T \rightarrow \infty$.  The calculated single-sided shape for two comb
lines, one in the center and one on the wing of the spectrum, is shown in Fig.~\ref{linewidth}a.
The dynamics of the noise on short time scales arising from the residual oscillatory part $\beta(T)$ of the second-order moments lead to features on the wings of the spectrum, as shown in Fig.~\ref{linewidth}a.
The most notable feature is the peak at 400 kHz,
the relaxation oscillation frequency.  The width of this peak depends on the 
strength of the effective saturable absorber that mode-locks the laser, which is 
strongly influenced by the pump power \cite{menyuk_pulse_2007} and can become 
overdamped under certain conditions.

The line is to a very good approximation Lorentzian.  
The FWHM linewidth of the $n^{\mathrm{th}}$ comb line is simply $\gamma_n = \gamma_{\theta\theta} - 2 \Omega_n \gamma_{\theta \tau} + \Omega_n^2 \gamma_{\tau\tau}$, where $\gamma_{xy}$ is the asymptotic slope of each corresponding term in the exponent of Eq.~(\ref{pow_spect_dens}).  The linewidth is plotted as a function of $\Omega_n$ in Fig.~\ref{linewidth}b.
The noise in the
phase $\Dth$ is dominated by energy fluctuations coupling into phase noise through the 
Kerr effect.  The Schawlow-Townes linewidth, the linewidth if the phase noise is 
driven by $D_{\theta\theta}$ alone, is more than two orders of magnitude smaller.  The
noise in the time $\Dt$ is dominated by fluctuations in the central frequency
coupling into timing jitter through cavity dispersion.  These central frequency 
fluctuations are driven mostly by the direct term $D_{\om\om}$.  The noise is dominated by 
timing jitter over most of the spectrum,
suggesting the importance of minimizing the cavity dispersion for comb 
applications.  Also note that the correlation term $\langle \Dth \Dt \rangle$
leads to a shift in the frequency of minimum linewidth from the central
frequency.  This shift is especially important for the Kerr effect, which affects both 
phase and timing \cite{ablowitz_carrier-envelope_2004}.  

\section{Feedback, the phase noise spectrum, and frequency uncertainty}
In metrological applications of femtosecond frequency 
combs, the comb is phase-locked to a frequency standard, and this has a
profound effect on the lineshape.  The moments involving only $\tau$ and
$\theta$ are now damped, and their time domain solutions now take the form $\alpha+\eta(T)$,
where $\alpha$ is a constant and $\eta(T)\rightarrow 0$ as $T\rightarrow\infty$.  
The first term leads to a delta function in frequency with an area proportional to $\exp (-\alpha)$.
This part of the comb line is coherent, which makes it useful in metrology applications.
The second term corresponds to residual phase noise, the shape of which depends on
the laser dynamics and the details of the feedback system.  In metrology applications, the residual phase noise
leads to uncertainty in frequency measurements \cite{nist_1337}.

In optical frequency metrology \cite{cundiff_2003} or optical clock development and comparison \cite{rosenband_2008,ludlow_2008}, $\fceo$ is measured and locked using self-referencing and $\frep$ is locked to a microwave frequency standard or to a heterodyne beat between a comb line and a laser line locked to an optical atomic transition.
Because it depends on the phase with respect to the envelope, $\fceo$ depends on 
both $\tau$ and $\theta$.  In general, the phase of a comb line depends on both parameters
as well, so each feedback system, for both $\fceo$ and $\frep$, will measure and act
on fluctuations in both $\tau$ and $\theta$.
If, for simplicity, one assumes that the feedback mechanism --- a cavity mirror on a piezoelectric transducer or an acousto-optic modulator controlling the pump power --- acts on $\Delta \fceo$ and $\Delta f_n$,
rather than some other linear combination of $\Delta \tau$ and $\Delta \theta$, one can proceed by defining a linear transformation which goes from the $\{\tau,\theta\}$ basis to the $\{\fceo,f_n\}$ basis.  Each feedback loop acts on the latter pair of variables, and the inverse transformation determines how the feedback affects $\tau$ and $\theta$.
We have modeled a feedback system with parallel proportional and integrator
stages because it is simple enough that analytical solutions can still be found.  The parameters we choose give a servo bump at about 1 kHz.
We assume a simple radio frequency $\frep$ lock rather than a lock using an optical heterodyne beat.
We have found that shot noise in the detection of beats used for locking the laser
introduces a 
negligible amount of noise through the feedback loops because the 
bandwidth of a typical feedback system is approximately two orders of 
magnitude smaller than $\frep$.
We note that when the feedback bandwidth becomes 
comparable to the dynamical frequencies in the laser, the laser 
dynamics participate in the feedback mechanism \cite{matos_carrier-envelope_2006}.  That is not the case 
in the system that we considered.

\begin{figure}
\center{\includegraphics[width=10cm]{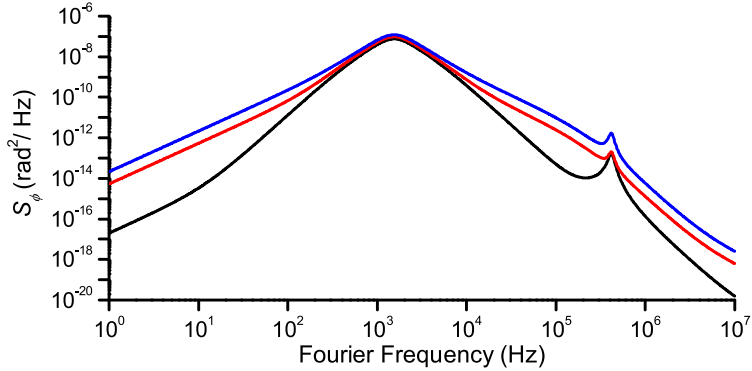}}
\caption{Calculated phase noise properties of the locked comb.  Phase noise spectrum for comb lines near $\omeq$ (black), 100 THz away (red), and 200 THz away (blue).}
\label{locked}
\end{figure}

For the $n^{\mathrm{th}}$ comb line, the phase noise spectrum is
\begin{equation}
S^\phi_n(\omega) = \frac{1}{2}\Omega_n^2 \langle[\Dt(\omega)]^2\rangle-\Omega_n\langle\Dt(\omega)\Dth(\omega)\rangle+\frac{1}{2}\langle[\Dth(\omega)]^2\rangle.
\label{phase_noise_spect_dens}
\end{equation}
We show this solution for several comb lines in Fig.~\ref{locked} for the given laser and feedback parameters.
While shot noise does not add much noise to the comb lines through the feedback, it does in general contribute strongly to the phase noise of a heterodyne beat, but we do not consider that here.
The phase noise spectrum contains all the information needed to determine the uncertainty in a frequency measurement.  
There has been recent work on how best to estimate uncertainty from measurements using modern frequency counters, which use interpolation to increase their accuracy over older style counters \cite{rubiola_on_2005,dawkins_considerations_2007}.  Counting of zero-crossings is well-described by the Allan variance, whereas the uncertainty in measurements using modern counters is better described by the \textit{triangle variance} \cite{dawkins_considerations_2007}.  We have calculated both uncertainty estimates using the SE-limited phase noise spectrum and found that, in the center of the spectrum where the noise is smallest, the Allan (triangle) deviation is 8$\times 10^{-18}$ (8$\times 10^{-21}$) for 1 second averaging time.  The large difference between the two arises from the triangle variance's stronger rejection of high frequency (compared to the gate time) noise.  For a comb line 200 THz away from the center of the spectrum, we find that the Allan (triangle) deviation is 1$\times 10^{-17}$ (1$\times 10^{-20}$) for 1 second averaging time.  Note that these values depend strongly on the feedback parameters.

\section{Conclusion}
Combs have already been shown to be comparable with 1-s Allan deviation on the order of 10$^{-17}$ fractional uncertainty \cite{ma_optical_2004}. The calculations presented here are of the same order of magnitude, but we caution that our calculation is specific to the laser studied in \cite{menyuk_pulse_2007} and \cite{wahlstrand_quantitative_2007}, using the simple feedback model we considered.
We note that our approach applies to technical
noise sources as well, if one adds them to the driving terms contained in $\sD$.  Using that information, coupled with experimentally-derived parameters for $\sA$ and a detailed and accurate model of the feedback system, it should be possible to make quantitative predictions of uncertainty in frequency measurements.  A program focused on optimizing the linear response of the laser system could be used to produce better laser designs for optical frequency metrology and other applications of femtosecond frequency combs.

We thank D. Howe, N. Newbury, and T. Schibli for useful discussions.  S.T.C. is a staff member in the NIST Quantum Physics Division.

\end{document}